\documentclass[%
twocolumn,
 longbibliography,
 amsmath,amssymb,
 aps,
 pre,
]{revtex4-1}

\usepackage{mathtools}
\usepackage{lipsum}
\usepackage{graphicx}
\usepackage{dcolumn}
\usepackage{bm}
\usepackage{comment}
\usepackage{afterpage}
\usepackage{xcolor}
\definecolor{sapphire}{HTML}{0067A5}
\usepackage[colorlinks=true,citecolor=sapphire,linkcolor=sapphire,urlcolor=sapphire]{hyperref}

\begin{document}
\title{ {The crumpling transition of active tethered membranes}} 
\author{M. C. Gandikota and A. Cacciuto}
\email{ac2822@columbia.edu}
\affiliation{Department of Chemistry, Columbia University\\ 3000 Broadway, New York, NY 10027\\ }

\begin{abstract}
\noindent We perform numerical simulations of active ideal and self-avoiding tethered membranes. Passive ideal membranes with bending interactions are known to exhibit a continuous crumpling transition between a low temperature flat phase and a high temperature crumpled phase. Conversely, self-avoiding membranes remain in an extended (flat) phase for all temperatures even in the absence of a bending energy. We find that the introduction of active fluctuations into the system produces a phase behavior that is overall consistent with that observed for passive membranes. The phases and the nature of the transition for ideal membranes is unchanged and active fluctuations can be remarkably accounted for by a simple rescaling of the temperature.
For the self-avoiding membrane, we find that the extended phase is preserved even in the presence of very large active fluctuations. 
\end{abstract}

\maketitle
\section{Introduction}
Membranes are two dimensional structures embedded in a higher dimensional space. They are ubiquitous in nature and they  have a fundamental role in the biology of the cell. They partition complex biochemical environments where different biological processes need to take place and regulate the intracellular traffic in eukaryotic cells~\cite{alberts_molecular_2008}. Biological membranes that are generated by self-assembly of phospholipids form a double-layered film a few nanometers in thickness. These amphiphilic molecules can freely diffuse on the membrane surface. In vitro, these membranes can be modeled as fluid films whose elastic properties are properly accounted for with the Helfrich free energy which includes a surface tension and a bending energy term~\cite{Helfrich}.

Their morphological and elastic properties in vivo depend, however, not only on thermal fluctuations, but also on the way the membranes interact with or are coupled to their environment. 
For instance, the membranes enveloping red blood cells are known to be strongly coupled to the cell's inner cytoskeletal 
structure,  the tethered nature of which, significantly impacts their elastic and morphological properties~\cite{Boal2012Jan,NelsonBook,Paraschiv2023}.
Furthermore, biological membranes are typically subject not just to thermal, but also to active fluctuations generated, by ATP consumption~\cite{turlier2016,betz2009atp,gnesotto2018broken,ben2011effective}, cascading biochemical reactions, pushing and pulling of actin filaments~\cite{alberts_molecular_2008}, 
and more in general they develop as a result of metabolic energy consumption in living cells.
 
Fluctuations of giant vesicles are also known to be enhanced by embedding a light-activated protein bacteriorhodopsin within the phospholipid bilayer~\cite{manneville1999activity}.
Active fluctuations are known to keep these systems highly dynamic and away from their equilibrium state  by energy dissipation  
that can produce strong non-equilibrium effects on the scale of the system~\cite{ramaswamy_mechanics_2010,marchetti_hydrodynamics_2013,zottl_emergent_2016}. 

In this paper, we are interested in understanding the role of active fluctuations on the behavior of elastic, tethered membranes. Unlike their fluid counterparts, the elements constituting tethered membranes are permanently linked to each other, and therefore resist shear strain and cannot flow~\cite{NelsonBook}.
 Beyond the actin–spectrin networks of red blood cell  cytoskeleton~\cite{Schmidt1993Feb,Lux2016Jan}, other examples of tethered membranes are: cross-polymerized membranes~\cite{Fendler1984}, gels~\cite{Georges2005Apr}, membranes made of close-packed nanoparticles~\cite{Mueggenburg2007Sep}, graphene and graphite-oxide sheets~\cite{wen1992crumpled,Stankovich2006Jul} polymer films~\cite{Huang2007Aug} and pollen grains~\cite{Katifori2010Apr}.

The behaviour of equilibrated tethered membranes is rather interesting~\cite{wiese2000,bowick2001}. In the absence of self-avoidance, thermal fluctuations play an important role in determining the equilibrium structure of tethered membranes. A continuous transition between a flat (extended) and a crumpled phase is observed depending on the relative strength of the thermal fluctuations and the bending energy~\cite{kantor1986}.  Curiously, self-avoiding tethered membranes, robustly remain in the flat phase at all temperatures~\cite{plischke1988,abraham1989,bowick1996flat}.
The stability of the flat phase is due to the non-linear coupling between the in-plane and out-of-plane modes of deformation of the membrane which renormalizes its bending rigidity in a way that the membrane becomes stiffer as its size increases~\cite{peliti1987,aronovitz1988,chaikin1995}. 
We refer the reader to reference~\cite{NelsonBook} for a comprehensive review of the physics of tethered membranes.
Crucially, for self-avoiding membranes in equilibrium, the flat phase is observed even in the absence of any bending rigidity.
Although the crumpled phase is not theoretically expected when the embedding dimension $d\leq d_c=4$~\cite{david1996,wiese1997}, the crumpling of the membrane is possible in higher dimensions as self-intersections become increasingly rare.
This result was tested using different numerical models, including a triangulated network of hard spheres imposing different degrees of self-avoidance~\cite{boal1989,abraham1990} and a plaquette model, where the hard spheres are removed and self-avoidance is imposed by explicitly preventing triangle-triangle intersections among any discrete parts of the membrane surface~\cite{bowick2001}. 
 
The role of active fluctuations on ideal and self-avoiding polymers has been studied intensely over the last decade, resulting in a variety of novel phenomenological behaviour~\cite{Loi2011Oct,Kaiser2014Jul,Harder2014Dec,Ghosh2014Sep,Shin2015Oct,Kaiser2015Mar,
Samanta2016Apr,Eisenstecken2016Aug,Chelakkot2014Mar,Isele-Holder2015Sep,Kaiser2015Mar,Bianco2018Nov,Harder2018Feb,Gandikota2022Mar,Das2022Mar,deviations2019,das2021,gandikota2023rectification}.
We also refer the reader to~\cite{Winkler2020Jul} for a recent review on active polymers.

More recently a significant effort has been put forward to understand the behavior of active fluid vesicles~\cite{Iyer2022Sep,Cagnetta2022Jan,Turlier2018Dec,Kulkarni2023Apr,Iyer2023Jan}, and  
it was recently suggested that the well known phenomenon of flickering observed in red-blood cells $-$ the tethered counterpart to fluid membranes $-$ breaks down the fluctuation-dissipation relation and can only be explained by the presence active, non-equilibrium  forces~\cite{Turlier2016May}. It is therefore important to understand the behavior of tethered membranes under the presence of active forces, not only for their biological relevance but also because of their possible applications in materials engineering.

 While the behavior of stiff self-avoiding membranes immersed in an active bath have been numerically investigated~\cite{mallory_anomalous_2015}, the problem of how the crumpling transition of ideal tethered membranes or the stability of the flat phase of the corresponding self-avoiding system is affected by active fluctuations still remains open. This paper addresses both questions. 

\section{Model and methods}
\begin{figure}[t]
	\centering
 \includegraphics[width=0.45\textwidth]{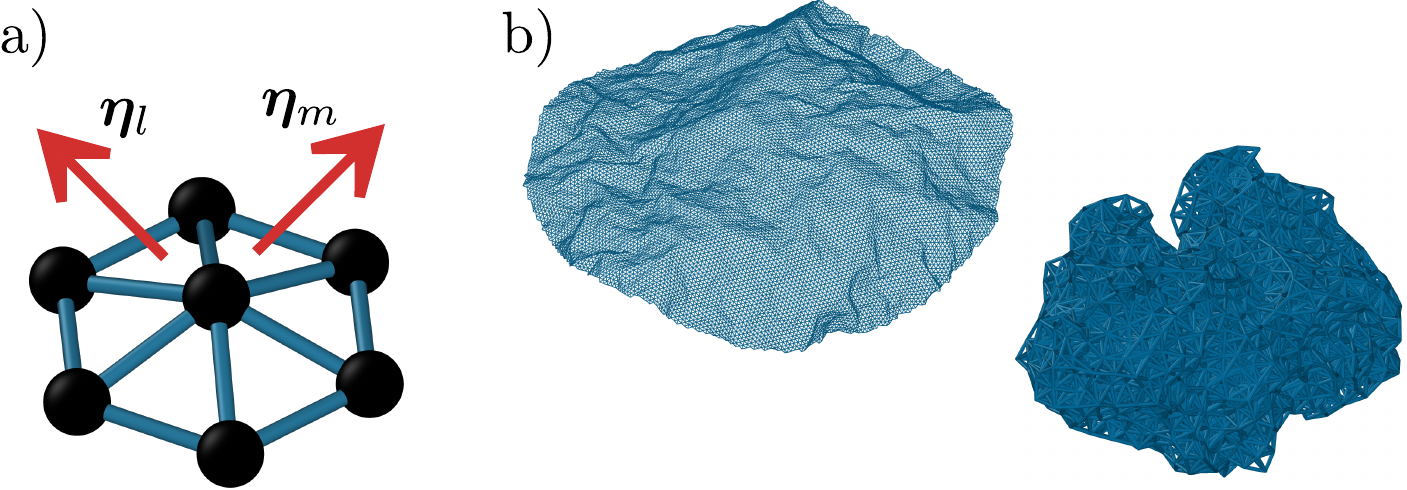}
	\caption{ a) A sketch of edge sharing triangular plaquettes with normals $\bm{\eta}_l,\bm{\eta}_m$. The spheres centered at the nodes represent self-avoiding particles of diameter $\sigma$ (reduced in size for visual clarity).  b) Simulation snapshots (spheres not shown) exhibiting the flat/extended phase and the crumpled phase of an ideal active tethered membrane. The crumpled phase snapshot is enlarged by 300\%. The system size is $N=13530$.}\label{membrane}
	\end{figure}

We model the elastic surface using a standard triangulated fishnet network representation~\cite{kantor_crumpling_1987} embedded in three dimensions. The rest shape of the membrane is circular, and apart from the boundary nodes, every inner node is six-coordinated. See Fig.~\ref{membrane} for a sketch of the membrane model and snapshots of the membrane in the flat and crumpled phase. 
To enforce self-avoiding interactions in a membrane with $N$ number of nodes, we place a spherical particle of diameter $\sigma$ at each node. Each of these node particles are connected to their nearest neighbors with a harmonic potential. 

If we denote the three dimensional coordinate of the $i^{\text{th}}$ node particle as $\bm{r}_i$, then the overall Hamiltonian of the system can be written as,
\begin{equation}\label{Hamiltonian}
\begin{split}
 H&=K\sum_{<ij>}(R_{ij}-\sigma)^2+\kappa\sum_{<lm>}(1-\bm{\eta}_l\cdot\bm{\eta}_m)\\
&+ 4\varepsilon\sum_{ij}\left[ \left( \frac{\sigma}{R_{ij}}\right)^{12} - \left(\frac{\sigma}{R_{ij}}\right)^{6} +\frac{1}{4}\right ],
\end{split}
\end{equation}
where $R_{ij}\equiv |\bm{r}_i-\bm{r}_j|$.
The first term accounts for the harmonic bonds between nearest neighbor particles and the equilibrium distance is set to $\sigma$.
For this membrane immersed in a solvent of temperature $T_0$, the spring constant $K=1500k_{\rm B}T_0$ is suitably set to a large value so that there is no appreciable stretching of the membrane. Here, $k_B$ is the Boltzmann constant. The second term is the bending energy where $\kappa$ is the bending constant and ($\bm{\eta}_l,\bm{\eta}_m$) represent the normal vectors of any two adjacent triangles (plaquettes)  sharing an edge (see Fig.~\ref{membrane} for a sketch of the model). The third term only applies to self-avoiding membranes, and models the excluded volume interaction between the node particles using the Weeks-Chandler-Andersen (WCA) potential cut off at $R_{ij}=2^{1/6}\sigma$ and set to zero beyond that distance. When considering ideal membranes, we set $\varepsilon=0$ $\forall R_{ij}$, otherwise we keep $\varepsilon=k_{\rm B}T_0$.

Activity is introduced in the system by adding a self-propelling force of constant magnitude $v_p$ to each of the node particles. The system dynamics is resolved using Brownian motion
according to
\begin{equation}\label{langevin}
\begin{split}
\frac{d\pmb{r}_i(t)}{dt}  &=  \frac{1}{\gamma} \pmb{f}_i +   v_p \,  \pmb{\hat{n}}_i(t)\,  + \sqrt{2D}\,\pmb{\xi}(t),\\
\frac{d \pmb{{\hat{n}}}_i(t) }{dt}&=\sqrt{2D_r}\, \pmb{\xi}_r(t) \times \pmb{\hat{n}}_i(t),
\end{split}
\end{equation}
where  $i$ is the particle index and the unit vector $\pmb{\hat{n}}$ is the axis of propulsion. The conservative forces on each particle are denoted by $\pmb{f}_i=-\partial H/\partial \bm{r}_i$. 
 The translational diffusion coefficient $D$, temperature $T_0$ and the translational friction $\gamma$ are constrained to follow the Stokes-Einstein relation $D=k_{\rm B}T_0\gamma^{-1}$. Likewise, the rotational diffusion coefficient is constrained to be $D_r=k_{\rm B}T_0\gamma_r^{-1}$, with $D_r = 3D\sigma^{-2}$. The Gaussian white-noise terms induced by the solvent for the translational $\pmb{\xi}$ and rotational $\pmb{\xi}_r$ motions are characterized by the relations $\langle \pmb{\xi}(t)\rangle = 0$ and $\langle \xi_m(t) \xi_n(t^\prime)\rangle = \delta_{mn}\delta(t-t^\prime)$.

 We perform molecular dynamics (MD) simulations using the numerical package LAMMPS~\cite{plimpton_fast_1995} and the units of length, time  and energy respectively are set to be $\sigma$, $\tau=\sigma^2D^{-1}$ and $k_{\rm B}T_0$. All simulations were run with a time step less than $\Delta t=2\times10^{-5}\tau$. 
 The strength of the active forces is quantified by the P\'eclet number defined as $Pe=v_p\sigma/D$.

\begin{figure*}[t]
	\centering
	\includegraphics[width=0.8\textwidth]{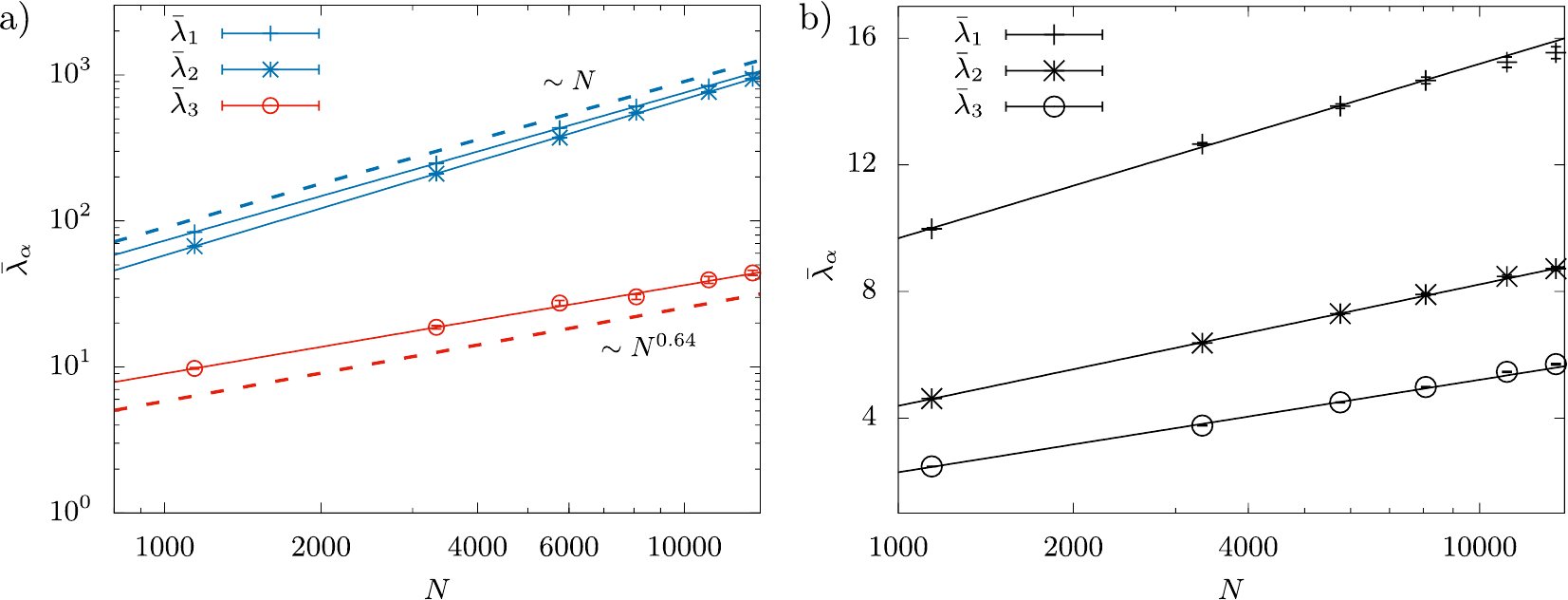}
	\caption{(a) Size dependence of the eigenvalues of the shape tensor for an ideal active tethered membrane in the flat phase at
 $Pe=20$ and $\kappa=10$.
 The largest two eigenvalues of the shape-tensor scale linearly with the size of the membrane, $N$, which confirms the flat phase of the ideal membrane. The smallest eigenvalue is compatible with a roughness exponent of 0.64. The dashed lines are the reference power laws for the passive system. (b) Logarithmic size dependence of the eigenvalues of the shape tensor for an ideal active membrane in the crumpled phase at $Pe=100$ and $\kappa=0$. The plot is in a Log-Linear scale. 
   The solid lines are fits to the data, and $\bar{\lambda}_{\alpha}\equiv\lambda_{\alpha}-c_{\alpha}$, where $c_{\alpha}$ is an additive constant included to the power law fits.}\label{ideal_scale}
\end{figure*}
The shape of the fluctuating membrane is assessed using the shape tensor~\cite{rudnick_aspherity_1986}, 
\begin{equation}
S_{\alpha\beta}=\frac{1}{2N^2}\sum_{i=1}^N\sum_{j=1}^N(r_{i\alpha}-r_{j \alpha})(r_{i \beta}-r_{j \beta}),
\end{equation}
where $\alpha,\beta=\{x,y,z\}$ (the three Cartesian coordinates) and $i,j$ are the particle indices. The trace of the shape tensor is the square of the radius of gyration. 
This symmetric tensor is diagonalized into three principal directions with corresponding eigenvalues $\lambda_\alpha$ where $\alpha={1,2,3}$ such that $\lambda_1 \geq \lambda_2 \geq \lambda_3$. The scaling of the time-averaged eigenvalues with membrane size,  $\left<\lambda_\alpha\right>\sim N^{\beta_\alpha}$, determines the phase of the membrane. 
  
For ideal membranes, the flat phase is characterized by a linear growth of the two largest eigenvalues with its overall size $N$. This implies $\beta_1=\beta_2=1$, while the third eigenvalue is associated with the roughness exponent $\zeta$, which accounts for the height fluctuations $h$ of the membrane and scales as 
$\left< h^2\right>=N^{\zeta}$, where $\zeta\simeq 0.64(4)$. 
In the isotropic crumpled phase, all three eigenvalues have a logarithmic dependence on $N$~\cite{kantor1986}.

Self-avoiding membranes lack a crumpling transition and are found in an extended state for all temperatures.
Its extended phase is characterized by exponents $\beta_{\alpha}$ compatible with those associated with the flat phase of the ideal membrane~\cite{bowick_statistical_2001}.
Adding attractive forces between the node particles on a self-avoiding membrane lead to the formation of a 
compact/folded phase with size exponent $\beta_{\alpha}=2/3\,\,,\forall\alpha$~\cite{abraham1990}. 

To estimate the size exponents in our numerical simulations, we fit the average eigenvalues of the shape tensor over six system sizes in the range of $N\in\{1142,13530\}$. We start with a flat initial configuration of the tethered membrane and allow $10^8$ MD steps for the system to reach the steady state. Configurations of the membrane are extracted every $10^5$ MD steps and the averaging  of the eigenvalues is done over a minimum of 3000 configurations for the self-avoiding case and a minimum of 6000 configurations for the ideal tethered membranes. The standard errors of the eigenvalues about their averages are used as the error bars for our data. To extract the shape exponents $\beta_\alpha$, we perform the least squares fit to the eigenvalues using a power law with and without an additive constant i.e. we fit $\lambda_\alpha=b_\alpha N^{\beta_\alpha}+c_\alpha$ and $\lambda_\alpha=b_\alpha N^{\beta_\alpha}$, as well as including a logarithmic correction $\lambda_\alpha=b_\alpha N^{\beta_\alpha}+c_\alpha\;\log(N)$.  This is done to control for possible sub-leading corrections~\cite{bowick2001}. The error on the exponents accounts for the range of values obtained with this fitting procedure.
 
\section{Ideal membranes}
We first consider the case of ideal membranes, i.e. tethered networks with no self-avoiding interactions ($\epsilon=0$ in Eq. \ref{Hamiltonian}).
Since the spring constant of the bonds is very large, effectively making the membrane unstretchable, the only relevant elastic parameter controlling the shape of the surface is the bending constant $\kappa$. In the competition between  bending energy which favors a flat phase and entropy which favors a crumpled phase,  ideal membranes are found in a flat phase at low temperatures and collapse into a crumpled phase at high temperatures. The phase transition is continuous and for a fixed value of the bending constant $\kappa$, the critical point occurs when $\kappa/(k_BT_c)=0.33$~\cite{NelsonBook}, here $T_c$ is the critical temperature.

We study the effect of active fluctuations on the phases of the membrane, the location of the transition point and the nature of the transition itself. For a given set of bending constants $\kappa$, this is done by studying the  conformational properties of the membrane as a function of P\'eclet number $Pe$.

\begin{table}[b]
\begin{center}
	\begin{tabular}[b]{|c|c|c|}
		\hline
		Exponents&Ideal & Self-avoiding\\
		\hline
		$\beta_1$&$ 1.01\pm0.03 $ &$0.91 \pm0.07 $\\
		$\beta_2$&$ 1.06\pm 0.04$ &$ 1.0\pm 0.1 $\\
		$~~\;\;\;\beta_3\; (\zeta)$&$ 0.6\pm 0.1$ &$ 0.67\pm 0.09 $\\
		\hline
	\end{tabular}
 \end{center}
\caption{Shape exponents for ideal ($Pe=20, \kappa=10$) and self-avoiding ($Pe=100, \kappa=0$) active tethered membranes.}\label{exp_tabl}
\end{table}

\begin{figure*}[t]
	\centering
	\includegraphics[width=0.8\textwidth]{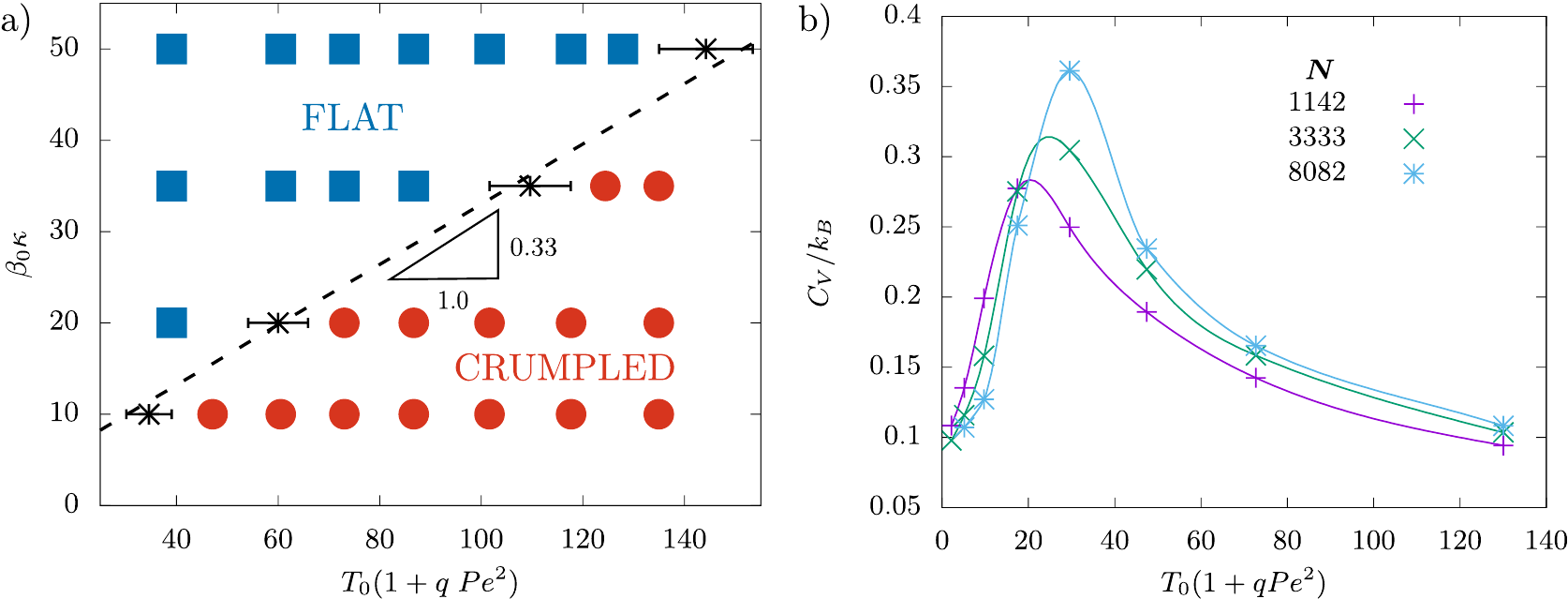}
	\caption{a) Phase diagram of ideal membranes for different values of bending rigidity and effective active temperature $T^{\text{eff}}=T_0(1+q Pe^2)$. Here, $\beta_0=1/(k_{\rm B}T_0)$ and $T_0$ refers to the background temperature of the thermal fluctuations. The phase of the membrane with a bending constant $\kappa$ is taken to be flat (blue squares)/crumpled (red disks) if the radius of gyration is greater/less than the critical radius of gyration at the same $\kappa$. The dashed line indicates the boundary between the two phases and follows the slope  $\kappa/(k_BT^{\text{eff}}_c)=0.33$ when an effective active temperature is defined with $q=1/42$. The system size is $N=3333$. 
 (b) Typical behavior of the effective specific heat, defined as $dE/dT^{\text{eff}}$, as a function of $T^{\text{eff}}$ for different values of membrane size $N$.} 
 \label{specific_heat}
	\end{figure*}
 
Our results show that for small values of $Pe$, the membrane retains its flat shape as depicted in Fig.~\ref{membrane} (r.h.s) and is characterized by shape exponents $\beta_1=\beta_2\simeq 1$ and a roughness exponent $\zeta\simeq 0.6$, within error bar of the value expected for the parent equilibrium system. All exponents, obtained by fitting the eigenvalues of the shape tensor to a power law of the form $\lambda_{\alpha}(N)=c_{\alpha}+d_{\alpha} N^{\beta_{\alpha}}$, are reported in Table~\ref{exp_tabl}. In Fig. \ref{ideal_scale}(a), we see the linear dependence of the two largest eigenvalues $\lambda_1,\lambda_2$ $-$ a signature of the flat phase $-$ as well as the dependence of the smallest eigenvalue $\lambda_3$ on the size of the membrane $N$.
For sufficiently large values of $Pe$, the ideal active membrane assumes a crumpled configurations as depicted in the bottom-left of 
 Fig.~\ref{membrane}. In this case, the membrane is isotropic and all eigenvalues scale logarithmically with $N$ as shown in Fig.~\ref{ideal_scale}(b) which depicts in a Log-Linear scale the behavior of $\lambda_{\alpha}$ as a function of $N$. Again, this phase is consistent with that observed for equilibrium ideal membranes. Thus, the character of the two equilibrium phases for ideal membranes is preserved under the influence of the active fluctuations. 

This nontrivial result suggests that the phase behavior of an active membrane may be encapsulated as an effective higher temperature of the system. Interestingly, while this mapping has been possible for some active problems, such as  the sedimentation of active colloidal suspensions under the influence of gravity~\cite{palacci2010} or the structure of flexible and semiflexible active polymers within the framework of dry active matter~\cite{Loi2011Oct,Kaiser2015Mar,CacciutoDas2021,Gandikota2022Mar}, there are many counterexamples where such mapping is completely inadequate; the most important one is probably the motility induced phase separation in a dense fluid of active particles~\cite{cates_motility-induced_2015}. Understanding the conditions  under which this mapping can be made for arbitrary active systems still remains an open question.

To test the feasibility of this mapping for ideal tethered membranes, we consider the standard functional form for the effective temperature $T^{\rm eff}=T_0(1+q\,Pe^2)$, where $q$ is a system dependent parameter. We then
compute the specific heat per particle, defined as $C_V\equiv dE/dT^{\rm eff}$ as a function of $T^{\rm eff}$. Here $E$ is the average energy per particle computed from Eq.~\ref{Hamiltonian}.
The peak of $C_V$ as a function of $T^{\rm eff}$ identifies the transition point, $T_c^{\rm eff}$.
By varying the only unknown parameter $q$, we check that membranes with different values of the bending constant $\kappa={10,20,35,50}$ across a range of P\'eclet numbers lead to the same transition point characterized by the ratio $\kappa/(k_{\rm B}T_c^{\rm eff})=0.33$ (the passive critical point at room temperature) for every value of $\kappa$. The results of this analysis can be seen in Fig.~\ref{specific_heat}(a), where we plot the rescaled bending constant $\kappa$ as a function of $T_c^{\rm eff}$. Strikingly, when we select $q\approx 1/42$, all points follow the same line with slope 0.33, making for a strong case that the physical properties of an active ideal tethered membrane can be properly described by its parent passive system with an effective higher temperature of $T^{\rm eff} \approx T_0(1+Pe^2/42)$.
Figure~\ref{specific_heat}(b) shows the divergence of $C_V$ as a function of $T^{\rm eff}$ at a given bending constant and different membrane sizes $N$, indicating the second order nature of the transition.

\section{Self-avoiding membranes}
As discussed above, the main difference between ideal and self-avoiding membranes is that the latter lack the crumpled to flat transition and tend to remain in an extended state even when the bending constant is set to zero. 
While it was argued that the self-avoidance of the hard balls in the tethered membranes impose a restriction on the range of the bending angles effectively yielding a small finite bending rigidity~\cite{abraham1990}, it was shown that even the use of very small spheres on top of the nodes of the lattice~\cite{boal1989} or enforcing excluded volume with just the second-nearest-neighbors~\cite{abraham1990}, rather than against all membrane particles, is sufficient for tethered membranes to lose the crumpling transition. Furthermore, simulations with the more computationally expensive  impenetrable plaquette model have also indicated the stability of the flat phase~\cite{bowick2001} for $\kappa=0$, suggesting that passive self-avoiding membranes retain their extended phase for all temperatures~\cite{plischke1988,abraham1989}. 

\begin{figure}[t]
	\centering
	\includegraphics[width=0.4\textwidth]{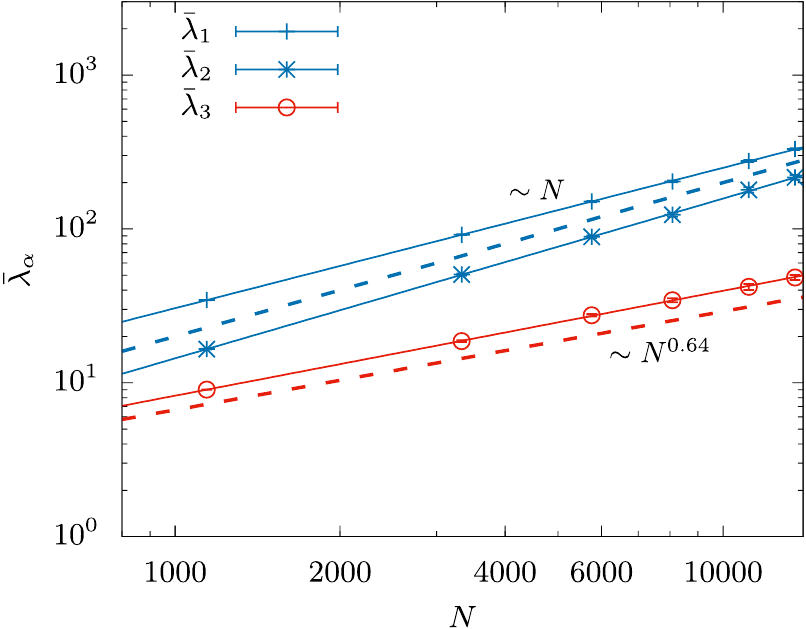}
	\caption{Size dependence of the eigenvalues of the shape tensor for an active tethered self-avoiding membrane. The largest two eigenvalues of the shape-tensor scale linearly with the size of the membrane, $N$, which confirms the flat phase of the self-avoiding membrane. The smallest eigenvalue is compatible with a roughness exponent of 0.64. In this plot $Pe=100$ and  $\kappa=0$. The dashed lines are the reference power laws for the passive system.  The solid lines are fits to the data, and $\bar{\lambda}_{\alpha}\equiv\lambda_{\alpha}-c_{\alpha}$, where $c_{\alpha}$ is an additive constant included to the power law fits.}
	\end{figure}
  
We therefore focus our attention to the specific case of $\kappa=0$, and we use the computationally efficient spring-and-balls model discussed above, where we enforce self-avoidance by imposing the excluded volume among all node particles on the membrane with diameter $\sigma$ as discussed in Eq.~\ref{Hamiltonian}. This choice is justified because the strength of active fluctuations is much larger than the effective bending rigidity, which was estimated to be of the order of $1.13\; k_{\rm B}T$~\cite{abraham1990}, intrinsic to the model. 

\begin{figure*}[btp]
	\centering
	\includegraphics[width=0.8\textwidth]{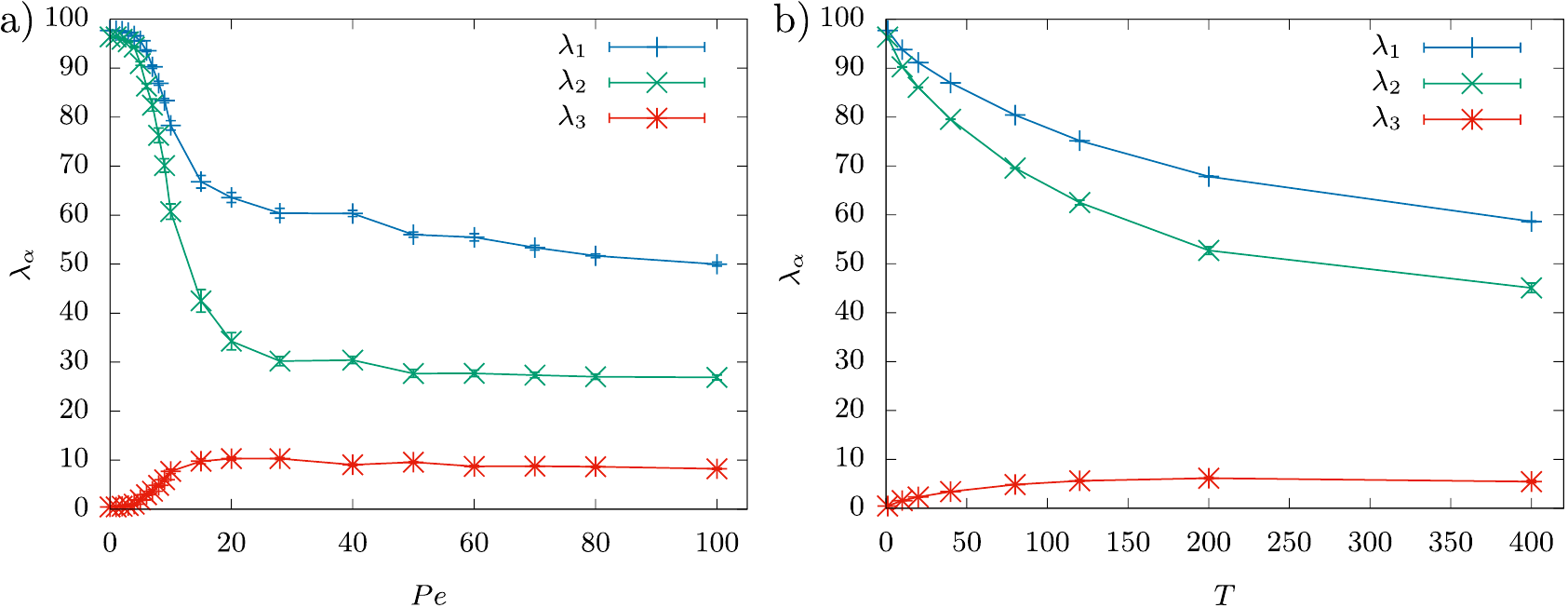}
	\caption{(a) Eigenvalues of shape-tensor, $\lambda_\alpha$, for active  self-avoiding membranes as a function of P\'eclet number $Pe$.
 (b) Eigenvalues of shape-tensor, $\lambda_\alpha$, for passive self-avoiding membranes as a function of temperature $T$. 
 In both cases the membrane was built with $N=1142$ nodes, the eigenvalues are sorted by size with $\lambda_1>\lambda_2>\lambda_3$, and the bending constant $\kappa$ is set to zero. }\label{dependence}
	\end{figure*} 
 
It should be stressed that although an effective temperature can properly describe the active fluctuations of an ideal membrane, this may not be necessarily true for the self-avoiding one. Yet, we find that upon turning on the activity, the self-avoiding membrane retains its extended phase for the broad range of values of $Pe\in{[0,100]}$ considered in this study. This is again established by studying the size dependence of the eigenvalues of the shape tensor. The corresponding exponents are shown in Table.~\ref{exp_tabl}, and indicate that the ones associated with the two largest eigenvalues are close to unity and the smallest one has a value compatible with the roughness exponent 0.6, suggesting  that the self-avoiding membrane retains its flat phase under active fluctuations and that this phase is consistent with that observed for the flat phase of the ideal membrane for $Pe<Pe_c$. 

It is also instructive to measure how the three eigenvalues of the shape tensor depend on the strength of the active forces, $Pe$. To this end, we performed a series of simulations of an active  self-avoiding membrane for different $Pe$. The results are shown in  Fig.~\ref{dependence}(a). 
The overall trend is that for small values of $Pe$ we observe a sharp decrease of the two largest eigenvalues of the shape tensor, accompanied by a corresponding sharp increase of the smallest eigenvalue. 
This behaviour is the result of the coupling between the in- and out-of-plane modes of deformation of the membrane. Upon further increasing  the strength of the active fluctuations the curves for all eigenvalues tend to plateau and overall remain very weakly dependent on  $Pe$. 
It should be stressed that the thickness of the membrane, associated with $\lambda_3$, always remains visibly smaller than the in-plane extent of the membrane associated with the two largest eigenvalues $\lambda_{1,2}$. This result reinforces that the membrane does not collapse and crumple even for the largest P\'eclet numbers, $Pe=100$, considered in this study. 

As a reference, we also computed the same plot as a function of temperature, $T$, for the passive self-avoiding membrane. The results are shown in Fig.~\ref{dependence}(b). The overall behavior is quite similar, with the only notable difference in the value of $\lambda_2$, and $\lambda_3$, in the large $Pe$ and $T$ limits. Specifically, $\lambda_2$ for the active systems, is almost twice as small as that for its passive counterpart, whereas the size of the out-of-plane fluctuations, $\lambda_3$, for the active membrane are roughly twice the size of those of the passive membrane.
A visual inspection of the membrane trajectories suggests that this is due to much larger fluctuations of the  edges for the active membranes, where partially folded conformations are also observed together with the typical flat ones. This leads to an overall smaller value of the second eigenvalue and correspondingly to a larger value of $\lambda_3$.

An analogous  plot for the ideal system in reported in the appendix\ref{appendix}. Even in this case we note that the crumpled phase of the passive membrane is smaller than that of the active membrane. Furthermore, the crumpling transition of the active ideal membrane is smoother compared to its passive counterpart.

\begin{figure}[t]
	\centering
	\includegraphics[width=0.45\textwidth]{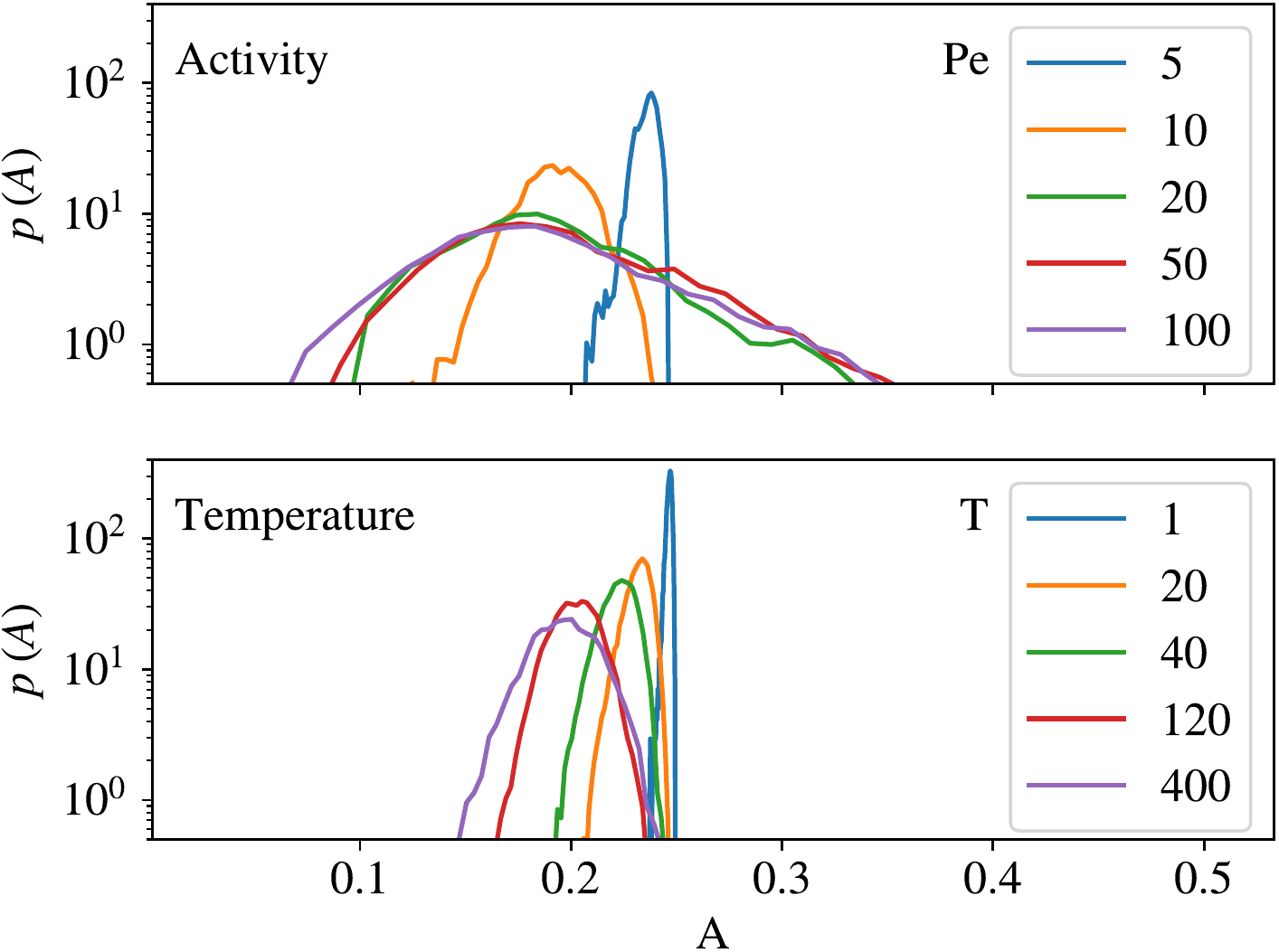}
	\caption{Probability distribution of the asphericity, $A$, for an active self-avoiding tethered membrane for different P\'eclet numbers $Pe$ (top) and for a passive self-avoiding tethered membrane for different temperatures, $T$ (bottom)}
 \label{asphericity}
\end{figure}

 To gain more insight into the shape fluctuations of these systems we also considered the distributions of the asphericity parameter for both active and passive membranes.
Following~\cite{rudnick_aspherity_1986}, we define asphericity as 
\begin{equation}
    A=\frac{3}{2}\frac{\lambda_1^2+\lambda_2^2+\lambda_3^2}{\left(\lambda_1+\lambda_2+\lambda_3\right)^2}-\frac{1}{2}.
\end{equation}
As a reference, this parameter is equal to zero for a fully isotropic surface ($\lambda_1=\lambda_2=\lambda_3$), it is equal to 1/4 for a perfectly flat surface ($\lambda_1=\lambda_2$ and $\lambda_3=0$), and $A\rightarrow 1$ for a linear/rod-like object corresponding to $\lambda_2=\lambda_3=0$.
Figure~\ref{asphericity} shows the results of this analysis for active (top) and passive (bottom) tethered surfaces. Again, superficially the trend is quite similar. Upon increasing the strength of the active forces (or the temperature of the passive system), the average asphericity decreases from around 1/4  for low active/thermal forces, and tend to plateau to a slightly smaller value of 0.18 for the active  and 0.20 for the passive membrane upon increasing the strength of the fluctuations. It is also quite evident that the distribution of the fluctuations for the active system is quite wider, confirming that the active membrane is capable of exploring shape fluctuations that are much more isotropic and much more anisotropic than its passive counterpart. Crucially, despite these differences, the size exponents for the active membrane point to an overall flat phase.

\section{Conclusions}
We performed numerical simulations of ideal and self-avoiding tethered membranes in the presence of active fluctuations. For simplicity, we used the standard active Brownian particles whose direction vectors are free to rotationally diffuse.

We find that the flat and crumpled phases, characteristic of passive ideal membranes, are retained  in the presence of active fluctuations. 
Remarkably, the role of active fluctuations on the transition point of the ideal membrane can be simply mapped to an equilibrium system with an effective temperature of the form $T^{\text{eff}}=T_0(1+q Pe^2)$ with $q=1/42$. 
Furthermore, the nature of the crumpling transition remains continuous, as suggested by a diverging specific heat at the critical effective temperature.

Our results on self-avoiding membranes indicate that even in the presence of  large active fluctuations, $Pe\rightarrow 100$, these surfaces retain their flat phase just as passive self-avoiding membranes remain flat at all temperatures. This is despite the fact that active forces allow for a broader range of shape fluctuations when compared to their passive counterparts as visible from the distribution function of the surface asphericity. Therefore, our results suggest that the flat phase of self-avoiding tethered membranes is quite robust,  at least within the range of activities considered in this study, and the collapse of these surfaces can only be accessible in the presence of explicit attractive interactions between different regions of the surface as is the case for passive membranes~\cite{abraham1990,abraham1991folding}.
 
It is also important to point out how the implicit incorporation of the active fluctuations within the nodes of the membranes (the approach discussed in this paper) leads to significantly different results with respect to the case when active fluctuations are induced by the presence of an explicit fluid of active particles pushing on a passive surface. In the latter case, activity results in bi-stable folded-to-extended states~\cite{Mallory2015Jul}.  Difference in phenomenological behavior was also recently observed when measuring the effective forces between active polymers~\cite{Gandikota2022Mar} depending upon how active forces are introduced into the system.

An interesting future direction for the study of active elastic membranes  is to consider the case of closed spherical shells. In passive systems, we know that thermal fluctuations, when renormalized, act as a negative internal pressure which can crumple a  shell~\cite{kovsmrlj2017}. A recent study~\cite{agrawal2022active} has been performed on spherical shells using Monte Carlo simulations that explicitly break detailed balance to mimic the effect of out of equilibrium fluctuations. Crumpling of the shell for different degrees of ``activity" was observed. Nevertheless, it would be interesting to see how explicit active fluctuations would affect the behavior of such shells, and whether a mapping of the active fluctuation onto an effective pressure is also possible in this case. Work in this direction is already underway and will be published elsewhere.

While the goal of this paper is to numerically investigate how non equilibrium active forces can alter the phase behavior of two dimensional microscopic elastic sheets, our results are likely to qualitatively capture the behavior of an elastic sheet in a bath containing active particles, and could be in principle realized experimentally by laterally cross-linking a two-dimensional condensate of active agents. For instance, one could consider a two-dimensional condensate of active emulsion droplets tethered with DNA linkers. Recently this set up was used to form an active chain of droplets where every droplet can freely perform rotational diffusion despite the tethering~\cite{kumar2023}.  

\section*{Acknowledgements}
A.C. acknowledges financial support from the National Science Foundation under Grant No. DMR-2003444. 

\bibliography{rsc} 
\bibliographystyle{apsrev4-1} 

\section*{Appendix}\label{appendix}
\renewcommand{\thefigure}{A\arabic{figure}}
\setcounter{figure}{0}
\begin{figure*}[t]
	\centering
	\includegraphics[width=0.8\textwidth]{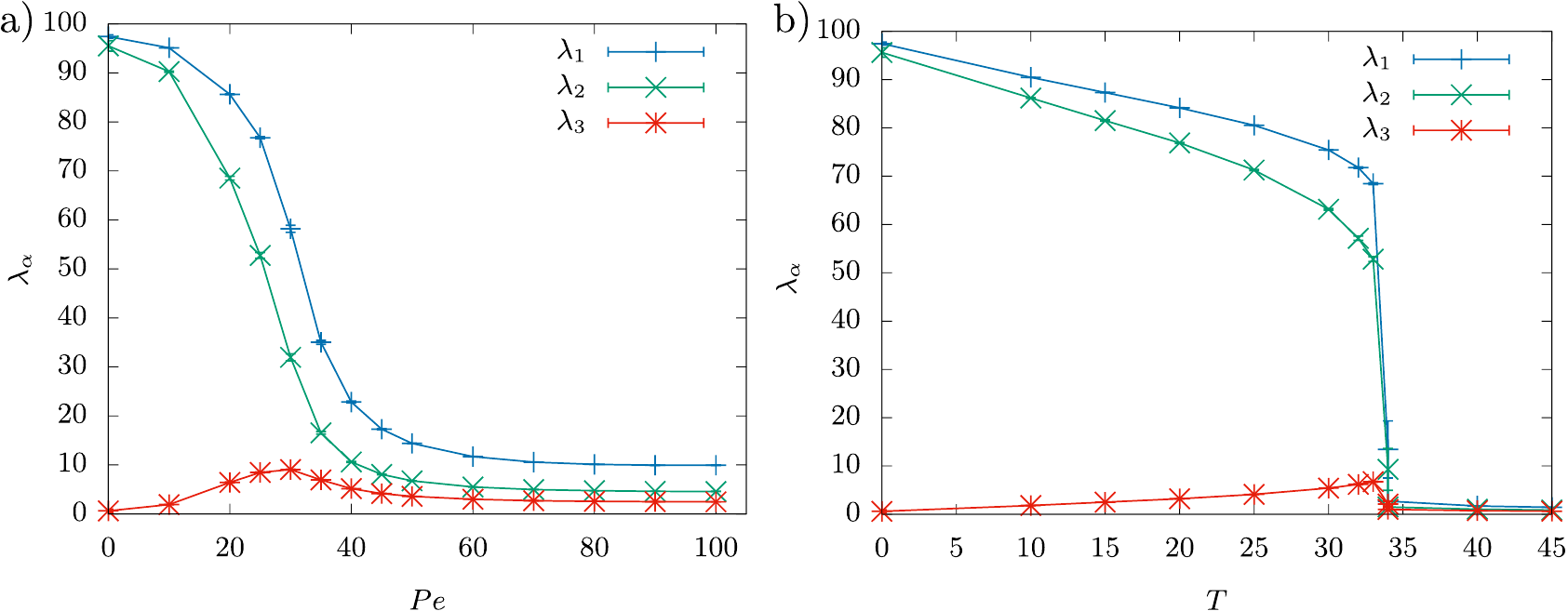}
	\caption{(a) Eigenvalues of shape-tensor, $\lambda_\alpha$, for  ideal membranes as a function of P\'eclet number $Pe$.
 (b) Eigenvalues of shape-tensor, $\lambda_\alpha$, for ideal membranes as a function of temperature $T$. 
 The system size of the membrane is $N=1142$ and the eigenvalues are sorted by size with $\lambda_1>\lambda_2>\lambda_3$. The bending constant $\kappa=10$. }
\end{figure*}

\end{document}